\documentclass[fleqn]{amsart}

\usepackage{amssymb}
\usepackage{amsfonts}
\usepackage[english]{babel}
\usepackage{euscript}
\usepackage{bbm}
\usepackage{xfrac}
\usepackage{color}

\theoremstyle{definition}

\theoremstyle{remark}

\numberwithin{equation}{section}

\newcommand{\fl}{\hspace*{-\mathindent}}
\newcommand{\textfrac}[2]{\textstyle{\frac{#1}{#2}}}

\begin{document}

\title[Lax representations via twisted extensions]{Lax representations via twisted extensions  of infinite-dimensional Lie algebras: 
some new results}

\author{Oleg I. Morozov}
\address{Trapeznikov 
Institute of Control Sciences,
  \\
65 Profsoyuznaya Street, Moscow 117997, Russia}
\email{oimorozov@gmail.com}


\date{April 1, 2022.% and, in revised form, June 22, 2001.  
}



\begin{abstract}
We find new integrable partial differential equations with Lax representations generated by extensions of Lie algebras of the 
Kac--Moody type as well as the Lie algebra of Hamiltonian vector fields on $\mathbb{R}^2$.
\end{abstract}

\maketitle

%======================================================================

\section{Introduction}

Lax representations provide the basic construction that allows applications of a num\-ber of techniques for studying nonlinear partial differential equations ({\sc pde}s),  whence   they are considered as the key feature indicating integrability thereof, see
\cite{WE,Zakharov82,RogersShadwick1982,NovikovManakovPitaevskiyZakharov1984,%
Konopelchenko1987,AblowitzClarkson1991,MatveevSalle1991,Olver1993,BacklundDarboux2001}
and references therein. Therefore  the problem of finding intrinsic conditions that ensure existence of a Lax representation for a given {\sc pde} is of great importance in the theory of integrable systems. In the recent papers \cite{Morozov2017} -- \cite{Morozov2021b} we propose an approach to tackle this problem. We have shown there that for a number of {\sc pde}s including the potential Khokhlov--Zabolotskaya equation, the Boyer--Finley equation, the hyper-CR equation of Einstein--Weyl structures, the reduced quasiclassical self-dual Yang--Mills equation, the 4D Mart{\'{\i}}nez Alonso--Shabat equation, the 4D universal hierarchy equation, 
and other equations, their known Lax representations can be inferred from non-triviality of the second twisted cohomology groups of the Lie algebras of contact symmetries of the {\sc pde}s. Moreover, we have shown that the technique allows one to find new Lax representations of some {\sc pde}s.

The aim of the present paper is to gain a better understanding of the relationship between the structure theory of infinite-dimensional 
Lie algebras and the theory of integrable systems. In particular, we construct new examples of deriving Lax representations of 
{\sc pde}s  from twisted extensions of some Lie algebras.    
      
In a number of above-mentioned  examples the symmetry algebras of the {\sc pde}s have  the form of the semi-direct sum 
$\mathfrak{s}_{\diamond} \ltimes \mathfrak{q}_{N,\varepsilon}$ of a finite-dimensional Lie algebra $\mathfrak{s}_{\diamond}$ 
and the infinite-dimensional Lie algebra $\mathfrak{q}_{N,\varepsilon}$  of the Kac--Moody type,
\cite{DavidKamranLeviWinternitz1985}, that is, the deformation of the tensor product 
$\mathfrak{q}_{N,0} = \mathbb{R}_N[s] \otimes \mathfrak{w}$, where $\mathbb{R}_N[s]$ is the commutative associative algebra 
of truncated polynomials of degree $N$ and  $\mathfrak{w}$ is the Lie algebra of vector fields on $\mathbb{R}$,  see Section \ref{section_3} for definition of $\mathfrak{q}_{N,\varepsilon}$. The second twisted cohomology groups of the Lie algebras 
$\mathfrak{s}_{\diamond} \ltimes \mathfrak{q}_{N,\varepsilon}$ from our examples turn out to be nontrivial, and the nontrivial
twisted 2-cocycles generate twisted extensions of these Lie algebras. Linear combinations of the Maurer--Cartan forms of the twisted extensions provide  the Wahlquist--Estabrook forms that generate Lax representations of the {\sc pde}s.

In examples from \cite{Morozov2017}--\cite{Morozov2021b}  we have $N \ge 3$, and the natural question is whether there exist Lie algebras of the form $\mathfrak{s}_{\diamond} \ltimes \mathfrak{q}_{N,\varepsilon}$ with $N<3$ and nontrivial second twisted cohomology groups whose twisted extensions generate Lax representations of some {\sc pde}s.     
In Sections \ref{section_4}, \ref{section_5}, and \ref{section_6} we present  three systems \eqref{first_eq} - \eqref{wt},  \eqref{HS2D} - \eqref{open_covering_2}, and \eqref{eq_08_04_2021} - \eqref{eq_08_04_2021_for_wt}  
whose Lax representations are generated by extensions of the Lie algebras  $\mathfrak{q}_{1,-1}$, $\mathfrak{q}_{1,-2}$, and $\mathfrak{q}_{2,-1}$, respectively. Equation \eqref{HS2D} can be considered as a 3D generalization of the generalized 2D Hunter--Saxton equation \cite{HunterSaxton1991,HunterZheng1994,BealsSattingerSzmigielski2001,Morozov2021c} with the special value of the parameter. We compare the structure of the symmetry algebras of the obtained systems with the structure of  the Lie algebras that generate the Lax representations. 

Paper \cite{Morozov2019} shows that for a number of known 4D integrable equations their Lax representations can be derived 
from the twisted extensions of the symmetry algebras, which turn out to be of the form 
$\mathfrak{s}_{\diamond} \ltimes (\mathbb{R}_N[s] \otimes \mathbb{R}[t] \otimes \mathfrak{w})$.
In  \cite{Morozov2021b} we apply the  technique  to a Lie algebra of this form and construct a new 4D integrable equation.
In the present paper we address an interesting question of finding new examples of integrable {\sc pde}s whose symmetry algebras include the Lie algebra $\mathfrak{h}$  of Hamiltonian vector fields on $\mathbb{R}^2$ as a subalgebra.  In Section \ref{section_7} we present such an example.  We derive equation \eqref{eq_5_02_2021} from  Lax representation defined by extensions of the Lie algebra  $\mathfrak{h} \oplus \mathfrak{w}$ . Other examples of integrable systems whose symmetry algebras include $\mathfrak{h}$ as `building blocks'  are given by the family of heavenly equations \cite{Plebanski1975,BoyerWinternitz1989,DoubrovFerapontov2010,KruglikovMorozov2012} and their `symmetric deformations' \cite{KruglikovMorozov2015}. Hence we refer equation \eqref{eq_5_02_2021} to as the `degenerate heavenly equation'.

%=====================================================================

\section{Preliminaries and notation}

\subsection{Symmetries and Lax representations}

The presentation in this section closely follows
\cite{KrasilshchikVerbovetsky2011}---\cite{KrasilshchikVinogradov1989} and  \cite{VK1999}.
All our considerations are local.
Let $\pi \colon \mathbb{R}^n \times \mathbb{R}^m \rightarrow \mathbb{R}^n$,
$\pi \colon (x^1, \dots, x^n, u^1, \dots, u^m) \mapsto (x^1, \dots, x^n)$, be a trivial bundle, and
$J^\infty(\pi)$ be the bundle of its jets of the infinite order. The local coordinates on $J^\infty(\pi)$ are
$(x^i,u^\alpha,u^\alpha_I)$, where $I=(i_1, \dots, i_n)$ are multi-indices with $i_k \ge 0$, and for every local section
$f \colon \mathbb{R}^n \rightarrow \mathbb{R}^n \times \mathbb{R}^m$ of $\pi$ the corresponding infinite jet
$j_\infty(f)$ is a section $j_\infty(f) \colon \mathbb{R}^n \rightarrow J^\infty(\pi)$ such that
$u^\alpha_I(j_\infty(f))
=\displaystyle{\frac{\partial ^{\#I} f^\alpha}{\partial x^I}}
=\displaystyle{\frac{\partial ^{i_1+\dots+i_n} f^\alpha}{(\partial x^1)^{i_1}\dots (\partial x^n)^{i_n}}}$.
We put $u^\alpha = u^\alpha_{(0,\dots,0)}$. Also, we will simplify notation in the following way: e.g., in the
case of $n=3$, $m=1$ we denote $x^1 = t$, $x^2= x$  $x^3= y$, 
and $u^1_{(i,j,k)}=u_{{t \dots t}{x \dots x}{y \dots y}}$ with $i$  times $t$, $j$  times $x$, and $k$ times $y$.

The  vector fields
\[
D_{x^k} = \frac{\partial}{\partial x^k} + \sum \limits_{\# I \ge 0} \sum \limits_{\alpha = 1}^m
u^\alpha_{I+1_{k}}\,\frac{\partial}{\partial u^\alpha_I},
\qquad k \in \{1,\dots,n\},
\]
$(i_1,\dots, i_k,\dots, i_n)+1_k = (i_1,\dots, i_k+1,\dots, i_n)$,  are called {\it total derivatives}.
They com\-mu\-te everywhere on
$J^\infty(\pi)$:  $[D_{x^i}, D_{x^j}] = 0$.

The {\it evolutionary vector field} associated to an arbitrary vector-valued smooth function
$\varphi \colon J^\infty(\pi) \rightarrow \mathbb{R}^m $ is the vector field
\[
\mathbf{E}_{\varphi} = \sum \limits_{\# I \ge 0} \sum \limits_{\alpha = 1}^m
D_I(\varphi^\alpha)\,\frac{\partial}{\partial u^\alpha_I}
\]
with $D_I=D_{(i_1,\dots\,i_n)} =D^{i_1}_{x^1} \circ \dots \circ D^{i_n}_{x^n}$.

A system of {\sc pde}s $F_r(x^i,u^\alpha_I) = 0$ of the order $s \ge 1$ with $\# I \le s$,
$r \in \{1,\dots, R\}$ for some $R \ge 1$,
defines the submanifold
$\EuScript{E}=\{(x^i,u^\alpha_I)\in J^\infty(\pi)\,\,\vert\,\,D_K(F_r(x^i,u^\alpha_I))=0,\,\,\# K\ge 0\}$
in $J^\infty(\pi)$.

A function $\varphi \colon J^\infty(\pi) \rightarrow \mathbb{R}^m$ is called a {\it (generator of an
infinitesimal) symmetry} of equation $\EuScript{E}$ when $\mathbf{E}_{\varphi}(F) = 0$ on $\EuScript{E}$. The
symmetry $\varphi$ is a solution to the {\it defining system}
\begin{equation}
\ell_{\EuScript{E}}(\varphi) = 0,
\label{defining_eqns}
\end{equation}
where $\ell_{\EuScript{E}} = \ell_F \vert_{\EuScript{E}}$ with the matrix differential operator
\[
\ell_F = \left(\sum \limits_{\# I \ge 0}\frac{\partial F_r}{\partial u^\alpha_I}\,D_I\right).
\]
The {\it symmetry algebra} $\mathrm{Sym} (\EuScript{E})$ of equation $\EuScript{E}$ is the linear space of
solutions to  (\ref{defining_eqns}) endowed with the structure of a Lie algebra over $\mathbb{R}$ by the
{\it Jacobi bracket} $\{\varphi,\psi\} = \mathbf{E}_{\varphi}(\psi) - \mathbf{E}_{\psi}(\varphi)$.
The {\it algebra of contact symmetries} $\mathrm{Sym}_0 (\EuScript{E})$ is the Lie subalgebra of $\mathrm{Sym} (\EuScript{E})$
defined as $\mathrm{Sym} (\EuScript{E}) \cap C^{\infty}(J^1(\pi))$. Symmetries with generators of the form 
$\varphi^\alpha = \eta^\alpha - \sum_i \xi^i\,u^\alpha_i$, $\eta^\alpha, \xi^i \in C^{\infty}(J^0(\pi))$, are referred to as
 {\it point symmetries}. They correspond  to vector fields 
$\sum_i \xi^i \,\partial_{x^i} + \sum_{\alpha} \eta^\alpha \, \partial_{u^\alpha}$ on $J^0(\pi)$.

Let the linear space $\EuScript{W}$ be either $\mathbb{R}^N$ for some $N \ge 1$ or  $\mathbb{R}^\infty$
endowed with  local co\-or\-di\-na\-tes $w^a$, $a \in \{1, \dots , N\}$ or  $a \in  \mathbb{N}$, respectively.
Variables $w^a$ are cal\-led {\it pseudopotentials} \cite{WE}.  Locally, a {\it differential covering} of $\EuScript{E}$ is 
a trivial bundle $\varpi \colon J^\infty(\pi) \times \EuScript{W} \rightarrow J^\infty(\pi)$ equipped with {\it extended total derivatives}
\[
\widetilde{D}_{x^k} = D_{x^k} + \sum \limits_{a}
T^a_k(x^i,u^\alpha_I,w^b)\,\frac{\partial }{\partial w^a}
\]
such that $[\widetilde{D}_{x^i}, \widetilde{D}_{x^j}]=0$ for all $i \not = j$ whenever $(x^i,u^\alpha_I) \in \EuScript{E}$. 
Define the partial derivatives of $w^a$ by  $w^s_{x^k} =  \widetilde{D}_{x^k}(w^s)$.  This yields the over-determined system
of {\sc pde}s 
\begin{equation}
w^a_{x^k} = T^a_k(x^i,u^\alpha_I,w^b)
\label{WE_prolongation_eqns}
\end{equation}
which is compatible whenever $(x^i,u^\alpha_I) \in \EuScript{E}$.
System \eqref{WE_prolongation_eqns} is referred to as the {\it covering equations}
or the {\it Lax representation} of equation $\EuScript{E}$.

Dually, the differential covering is defined by the
{\it Wahlquist--Estabrook forms}
\begin{equation}
\tau^a =d w^a - \sum \limits_{k=1}^{m} T^a_k(x^i,u^\alpha_I,w^b)\,dx^k
\label{WEfs}
\end{equation}
as follows: when $w^a$  and $u^\alpha$ are considered to be functions of $x^1$, ... , $x^n$, forms \eqref{WEfs}
are equal to zero if and only if  system \eqref{WE_prolongation_eqns} holds.

\subsection{Twisted cohomology of Lie algebras}

For a Lie algebra  $\mathfrak{g}$ over $\mathbb{R}$, its representation $\rho \colon \mathfrak{g} \rightarrow \mathrm{End}(V)$,
and $k \ge 1$ let $C^k(\mathfrak{g}, V) =\mathrm{Hom}(\Lambda^k(\mathfrak{g}), V)$ be the space of all $k$--linear 
skew-symmetric mappings from $\mathfrak{g}$ to $V$. Then the Chevalley--Eilenberg differential complex
\[
V=C^0(\mathfrak{g}, V) \stackrel{d}{\longrightarrow} C^1(\mathfrak{g}, V)
\stackrel{d}{\longrightarrow} \dots \stackrel{d}{\longrightarrow}
C^k(\mathfrak{g}, V) \stackrel{d}{\longrightarrow} C^{k+1}(\mathfrak{g}, V)
\stackrel{d}{\longrightarrow} \dots
\]
is generated by the differential $d \colon \theta \mapsto d\theta$ such that
\[
d \theta (X_1, ... , X_{k+1}) =
\sum\limits_{q=1}^{k+1} (-1)^{q+1} \rho (X_q)\,(\theta (X_1, ... ,\hat{X}_q, ... ,  X_{k+1}))
\]
\begin{equation}
\quad
+\sum\limits_{1\le p < q \le k+1} (-1)^{p+q+1}
\theta ([X_p,X_q],X_1, ... ,\hat{X}_p, ... ,\hat{X}_q, ... ,  X_{k+1}).
\label{CE_differential}
\end{equation}
The cohomology groups of the complex $(C^{*}(\mathfrak{g}, V), d)$ are referred to as the 
{\it cohomology groups of the Lie algebra} $\mathfrak{g}$ {\it with coefficients in the representation}  $\rho$. For the trivial representation $\rho_0 \colon \mathfrak{g} \rightarrow \mathbb{R}$, $\rho_0 \colon X \mapsto 0$, the cohomology groups are denoted by $H^{*}(\mathfrak{g})$.

Consider a Lie algebra $\mathfrak{g}$ over $\mathbb{R}$ with non-trivial first cohomology group $H^1(\mathfrak{g})$ and take 
1-form $\alpha \neq 0$ on $\mathfrak{g}$ such that $d\alpha =0$. Then for each $c \in \mathbb{R}$ define the 
{\it twisted differential}
$d_{c \alpha} \colon C^k(\mathfrak{g},\mathbb{R}) \rightarrow C^{k+1}(\mathfrak{g},\mathbb{R})$ by
the formula
\[
d_{c \alpha} \theta = d \theta - c \,\alpha \wedge \theta.
\]
From  $d\alpha = 0$ it follows that $d_{c \alpha} ^2=0$. The cohomology groups of the complex
\[
C^1(\mathfrak{g}, \mathbb{R})
\stackrel{d_{c \alpha}}{\longrightarrow}
\dots
\stackrel{d_{c \alpha}}{\longrightarrow}
C^k(\mathfrak{g}, \mathbb{R})
\stackrel{d_{c \alpha}}{\longrightarrow}
C^{k+1}(\mathfrak{g}, \mathbb{R})
\stackrel{d_{c \alpha}}{\longrightarrow} \dots
\]
are referred to as the {\it twisted} {\it cohomology groups}  \cite{Novikov2002,Novikov2005} of $\mathfrak{g}$
and denoted by $H^{*}_{c\alpha}(\mathfrak{g})$.

\section{Lie algebras of the Kac--Moody type and their extensions}
\label{section_3}

Consider the Lie algebra $\mathfrak{q}_{N,0} = \mathbb{R}_N[s] \otimes \mathfrak{w}$, where 
$\mathbb{R}_N[s] =  \mathbb{R}[s] / \langle s^{N+1} =0\rangle$ is the commutative unital algebra of truncated polynomials of variable $s$  of degree $N$, and $\mathfrak{w} = \langle V_k \,\,\vert\,\, k \ge 0 \rangle$,
$\displaystyle{V_k = \frac{1}{k!}\,t^{k}\,\partial_t}$, is the Lie algebra of polynomial vector fields on $\mathbb{R}$ referred to as 
the (one-sided) Witt algebra. Algebra $\mathfrak{q}_{N,0}$ admits the 
deformation\footnote[1]{For the full description of deformations of the Lie algebra $\mathfrak{q}_{N,0}$  see   \cite{Zusmanovich2003}.} 
generated by cocycle $\Psi \in H^2(\mathfrak{q}_{N,0}, \mathfrak{q}_{N,0})$, 
\[
\fl
\Psi (s^p \otimes V_m, s^q \otimes V_n) =
\left\{
\begin{array}{lcl}
\displaystyle{\frac{p\,n-q\,m}{m+n}}\,
{m+n\choose m}\,
s^{p+q} \otimes V_{m+n-1}, &&  m+n \ge 1, 
p+q \le N,
\\
0, && \mathrm{otherwise}.
\end{array}
\right.
\]
For each $\varepsilon \neq 0$ this cocycle defines new bracket
$[\cdot, \cdot]_{\varepsilon} = [\cdot, \cdot] +\varepsilon\,\Psi(\cdot, \cdot)$ on the linear space
$\langle s^p \otimes V_m \,\vert \, p \le N, m \ge 0 \rangle$. We denote the resulting Lie algebra as $\mathfrak{q}_{N,\varepsilon}$.  In other words, the Lie algebra $\mathfrak{q}_{N,\varepsilon}$ is isomorphic to the linear space of functions  
$f(t,s) = f_0(t)+s\,f_1(t)+\dots+s^N\,f_N(t)$,  $f_k\in \mathbb{R}[t]$,  equipped with the bracket
\begin{equation}
[f,g]_\varepsilon =f\,g_t-g\,f_t+\varepsilon\,s\,(f_s\,g_t-g_s\,f_t)
\label{epsilon_bracket}
\end{equation}
 such that there holds $s^k =0$ for $k > N$. Likewise to \cite{DavidKamranLeviWinternitz1985} it can be shown that 
$\mathfrak{q}_{N,\varepsilon} \subsetneq \mathfrak{g}(A_M^{(1)})$ for some $M \ge N$, see \cite{Kac1990} for definition of the Lie algebra $\mathfrak{g}(A_M^{(1)})$. Therefore $\mathfrak{q}_{N,\varepsilon}$ are referred to as Lie algebras of the Kac--Moody type.

Consider the dual 1-forms $\theta_{p,k}$ to the basis $s^q \otimes V_m$ of $\mathfrak{q}_{N,\varepsilon}$, that is, the linear
mappings $\theta_{p,k} \colon \mathfrak{q}_{N,\varepsilon} \rightarrow \mathbb{R}$ such that
$\theta_{p,k}(s^p\otimes V_m) = \delta_{p,q}\,\delta_{k,m}$. Define the formal series
\begin{equation}
\Theta = \sum \limits_{k=0}^{N} \sum \limits_{m=0}^{\infty} \frac{h_0^k h_1^m}{m!} \theta_{k,m}
\label{Theta_N_def}
\end{equation}
with formal parameters $h_0$ and $h_1$ such that $h_0^k =0$ when $k>N$ and $dh_0=dh_1=0$. Then \eqref{CE_differential} and \eqref{epsilon_bracket} entail the {\it structure equations}
\begin{equation}
d\Theta = \Theta_{h_1} \wedge (\Theta +\varepsilon \,h_0\,\Theta_{h_0})
\label{qNe_SE}
\end{equation}
of the Lie algebra $\mathfrak{q}_{N,\varepsilon}$. Here and below we use notion
$\Theta_{h_i} =\partial_{h_i} \Theta$ for partial de\-ri\-va\-ti\-ves of the formal series of 1-forms with respect to  the formal
pa\-ra\-me\-ters $h_i$.

For each $N \in \mathbb{N}$ and $\varepsilon \in \mathbb{R}$ the map 
$D_0 \colon s^p \otimes v_k \mapsto p\,s^p \otimes v_k$ is an outer  derivation of $\mathfrak{q}_{N,\varepsilon}$. 
Denote by $\mathfrak{a}_0\ltimes \mathfrak{q}_{N,\varepsilon}$ the associated one-dimensional `right' extension, 
\cite[\S 1.4.4]{Fuks1984}, of $\mathfrak{q}_{N,\varepsilon}$.  As a vector space
$\mathfrak{a}_0\ltimes \mathfrak{q}_{N,\varepsilon} = \langle Z_0 \rangle \oplus \mathfrak{q}_{N,\varepsilon}$,
and the bracket on $\mathfrak{q}_{N,\varepsilon}$ is extended to $Z_0$ by the formula
$[Z_0,s^p\otimes V_k]_{\varepsilon} = D_0(s^p\otimes V_k) = p\,s^p \otimes V_k$.

Let $\alpha_0 \colon \mathfrak{a}_0\ltimes \mathfrak{q}_{N,\varepsilon} \rightarrow \mathbb{R}$ be the dual form to $-Z_0$, that 
is, $\alpha_0(Z_0) =-1$ and $\alpha_0(s^p \otimes V_k) = 0$. Then the  structure equations for 
$\mathfrak{a}_0\ltimes \mathfrak{q}_{N,\varepsilon}$ read
\[
\left\{
\begin{array}{lcl}
d\Theta &=& \Theta_{h_1} \wedge (\Theta +\varepsilon \,h_0\,\Theta_{h_0})
 + h_0\,\alpha_0 \wedge \Theta_{h_0},
\\
d\alpha_0 &=&0.
\end{array}
\right.
\]

For some values of $N$ and $\varepsilon$ the Lie algebras $\mathfrak{s}_0\ltimes \mathfrak{q}_{N,\varepsilon}$ admit further
right extensions. In \cite{Morozov2017,Morozov2018,Morozov2019,Morozov2021a} we have shown examples of integrable
{\sc pde}s whose Lax representations can be inferred from such extensions when  $N\ge 3$. In the next three sections
we consider integrable {\sc pde}s that are related to extensions of $\mathfrak{a}_0\ltimes \mathfrak{q}_{1,-1}$,
$\mathfrak{a}_0\ltimes \mathfrak{q}_{1,-1/2}$, and $\mathfrak{a}_0\ltimes \mathfrak{q}_{1,-2}$.

As it was shown in \cite{Morozov2018}, for $N\in \mathbb{N}$ and $\varepsilon = -r^{-1}$ with
$r \in \{1, \dots, N\}$  the Lie algebra $\mathfrak{a}_0\ltimes \mathfrak{q}_{N,\varepsilon}$ admits the right extension
$\mathfrak{a}_1\ltimes \mathfrak{q}_{N,\varepsilon}$
generated by the outer derivation
$D_1 \colon \mathfrak{a}_0\ltimes \mathfrak{q}_{N,\varepsilon} \rightarrow \mathfrak{a}_0\ltimes \mathfrak{q}_{N,\varepsilon}$
with
\[
D_1 (s^p\otimes V_k) =
\left\{
\begin{array}{lcl}
k\,s^{p+r}\otimes V_{k-1}, && p+r \le N, k \ge 1,
\\
0, && \mathrm{otherwise}.
\end{array}
\right.
\]
We have $[D_0,D_1]=D_0 \circ D_1-D_1 \circ D_0=r\,D_1$. Then $\mathfrak{a}_1\ltimes \mathfrak{q}_{N,-1/r}$ as a vector space 
is $\langle Z_1\rangle \oplus (\mathfrak{a}_0\ltimes \mathfrak{q}_{N,-1/r})$, with the extension of the bracket of 
$\mathfrak{q}_{N,-1/r}$ given by $[Z_0,Z_1]_{-1/r} = -r\,Z_1$ and $[Z_1, s^p\otimes V_k]_{-1/r} = D_1(s^p\otimes V_k)$.
Consider the dual form $\alpha_1$ to the vector $-Z_1$, that, put $\alpha_1(Z_1)=-1$, 
$\alpha_1(Z_0)=\alpha_1(s^p\otimes V_k)=0$.  Then the structure equations for the Lie algebra 
$\mathfrak{a}_1\ltimes \mathfrak{q}_{N,-1/r}$   get the form
\begin{equation}
\left\{
\begin{array}{lcl}
d\Theta &=& \displaystyle{
\Theta_{h_1} \wedge \left(\Theta -\frac{h_0}{r} \,\Theta_{h_0}- h_0^r\,\alpha_1\right)
 + h_0\,\alpha_0 \wedge \Theta_{h_0},}
\\
d\alpha_0 &=&\displaystyle{0,\phantom{\frac{A}{A}}}
\\
d\alpha_1 &=& \displaystyle{r\,\alpha_0 \wedge \alpha_1.\phantom{\frac{A}{A}}}
\end{array}
\right.
\label{tilde_qNe_SE}
\end{equation}
These equations yield $H^1(\mathfrak{a}_1\ltimes \mathfrak{q}_{N,-1/r}) = \langle \alpha_0 \rangle$ and
$[ \alpha_0 \wedge \alpha_1 ] \in H^2_{r \alpha_0}(\mathfrak{a}_1\ltimes \mathfrak{q}_{N,-1/r})$,
therefore the Lie algebra $\mathfrak{a}_1\ltimes \mathfrak{q}_{N,-1/r}$ admits the twisted extension
$\mathfrak{a}_2\ltimes \mathfrak{q}_{N,-1/r}$ with the structure equations obtained by appending equation
\begin{equation}
d\alpha_2 = r\,\alpha_0 \wedge \alpha_2+\alpha_0 \wedge \alpha_1
\label{d_alpha_2_SE}
\end{equation}
to system \eqref{tilde_qNe_SE}.

Furthermore, we have
$[ \alpha_1 \wedge \alpha_2 ] \in H^2_{2 r \alpha_0}(\mathfrak{a}_2\ltimes \mathfrak{q}_{N,-1/r})$, therefore the Lie algebra
$\mathfrak{a}_2\ltimes \mathfrak{q}_{N,-1/r}$ admits the twisted extension $\mathfrak{a}_3\ltimes \mathfrak{q}_{N,-1/r}$
whose structure equations are obtained by appending equation
\begin{equation}
d\alpha_3 = 2\,r\,\alpha_0 \wedge \alpha_3+\alpha_1 \wedge \alpha_2
\label{d_alpha_3_SE}
\end{equation}
to system \eqref{tilde_qNe_SE}, \eqref{d_alpha_2_SE}. This process can be repeated: 
$\mathfrak{a}_k\ltimes \mathfrak{q}_{N,-1/r}$ admits the twisted extension 
$\mathfrak{a}_{k+1}\ltimes \mathfrak{q}_{N,-1/r}$ defined by the twisted 2-cocycle
$[ \alpha_1 \wedge \alpha_{k} ] \in H^2_{k r \alpha_0}(\mathfrak{a}_{k}\ltimes \mathfrak{q}_{N,-1/r})$.
In other words, the extended Lie algebra $\mathfrak{a}_{k+1}\ltimes \mathfrak{q}_{N,-1/r}$ has
the structure equations
\begin{equation}
\left\{
\begin{array}{lcl}
d\Theta &=& \displaystyle{
\Theta_{h_1} \wedge \left(\Theta -\frac{h_0}{r} \,\Theta_{h_0}- h_0^r\,\alpha_1\right)
 + h_0\,\alpha_0 \wedge \Theta_{h_0},}
 \\
d\alpha_0 &=&\displaystyle{0,\phantom{\frac{A}{A}}}
\\
d\alpha_1 &=& \displaystyle{r\,\alpha_0 \wedge \alpha_1,\phantom{\frac{A}{A}}}
\\
d\alpha_{m+1} &=& \displaystyle{m\,r\,\alpha_0 \wedge \alpha_{m+1}+\alpha_1 \wedge \alpha_m,
\qquad m \in \{1, \dots, k\}.
\phantom{\frac{A}{A}}}
\end{array}
\right.
\label{hat_qNek_SE}
\end{equation}

%=====================================================================

\section{Integrable equation associated to $\mathfrak{a}_{2}\ltimes \mathfrak{q}_{1,-1}$}
\label{section_4}

\vskip 7 pt
\noindent
Consider the Lie algebra $\mathfrak{a}_{3}\ltimes \mathfrak{q}_{1,-1}$ defined by the structure equations
\eqref{hat_qNek_SE} with $r=1$ and $k=2$, that is, by system
\begin{equation}
\left\{
\begin{array}{lcl}
d\Theta &=& \displaystyle{
\Theta_{h_1} \wedge \left(\Theta -h_0\,\Theta_{h_0}- h_0\,\alpha_1\right)
 + h_0\,\alpha_0 \wedge \Theta_{h_0},}
\\
d\alpha_0 &=&\displaystyle{0,\phantom{\frac{A}{A}}}
\\
d\alpha_1 &=& \displaystyle{\alpha_0 \wedge \alpha_1,\phantom{\frac{A}{A}}}
\\
d\alpha_2 &=& \displaystyle{\alpha_0 \wedge \alpha_2 +\alpha_0\wedge \alpha_1,\phantom{\frac{A}{A}}}
\\
d\alpha_3 &=& \displaystyle{2\,\alpha_0 \wedge \alpha_3 +\alpha_0\wedge \alpha_2.\phantom{\frac{A}{A}}}
\end{array}
\right.
\label{hat_q1-12_SE}
\end{equation}
Frobenius' theorem allows one to integrate equations \eqref{hat_q1-12_SE} step by step. In particular, we have
\[
\alpha_0 = \frac{dq}{q},
\quad
\alpha_1 =q\,dy,
\quad
\alpha_2 = q\,(dw +\ln q\,\,dy),
\quad
\alpha_3 = q^2\,(dv-w\,dy),
\]
\[
\theta_{0,0} = a\,dt,
\quad
\theta_{1,0} = q\,(dx - \ln a \, \,dy +u \,dt),
\quad
\theta_{1,1} =q\,a^{-1}\,(du - b\,dy -p\,dt),
\]
where $q \neq 0$, $a \neq 0$, $t$, $x$, $y$, $u$, $v$, $w$ are free parameters (‘constants of integration’). We
do not need explicit expressions for the other forms $\theta_{i,j}$ in what follows.

We proceed by imposing the requirement for the linear combination
$\theta_{1,1}-\theta_{1,0} = q\,a^{-1}\,(du -a\,dx - (b- a\,\ln a) \,dy -(p+a\,u)\,dt)$
to be a multiple of the contact form $du-u_t\,dt-u_x\,dx-u_y\,dy$ on the bundle of jets of sections of the bundle 
$\pi \colon \mathbb{R}^4 \rightarrow \mathbb{R}^3$, $\pi \colon  (t, x, y,  u) \mapsto (t, x, y)$,
that is, we take $a=u_x$, $b=u_y+u_x\,\ln u_x$, $p=u_t-u\,u_x$. Then we consider the linear combination
$\tau = \alpha_3-\theta_{1,0} = q^2 \,(dv- q^{-1}(u\,dt +dx+(q\,w-\ln u_x)\,dy)$ and put $q=v_x^{-1}$. This yields
\[
\tau = v_x^{-2} \,(dv- u\,v_x\,dt -v_x\,dx -(w-v_x\,\ln u_x)\,dy).
\]
The restriction of form  $\tau$ on the bundle of sections of the bundle $\mathbb{R}^6 \rightarrow \mathbb{R}^3$,
$(t,x,y,u,v,w) \mapsto (t,x,y)$,  gives the over-determined system of {\sc pde}s
\begin{equation}
\left\{
\begin{array}{lcl}
v_t &=& u\,v_x,
\\
v_y &=&w -v_x\,\ln u_x.
\end{array}
\right.
\label{open_covering_1}
\end{equation}
The integrability condition $(v_t)_y = (v_y)_t$ thereof gives two equations
\begin{equation}
u_{tx}=u\,u_{xx}-u_x^2\,\ln u_x-u_x\,u_y
\label{first_eq}
\end{equation}
and  
\begin{equation}
w_t = u\,w_x.
\label{wt}
\end{equation}
Thus system \eqref{open_covering_1} provides a Lax representation of system \eqref{first_eq}, \eqref{wt}. We can  find the Lax representation of equation \eqref{first_eq} by proceeding  as follows. Equation \eqref{wt} is a copy of the first equation in \eqref{open_covering_1}. We consider $w$ as a new pseudopotential and include equation \eqref{wt} into the Lax representation for equation \eqref{first_eq} by adding the copy $w_y = w_1 -w_x\,\ln u_x$ of the second equation from \eqref{open_covering_1} with additional function $w_1$, and then repeat this process. In other words, we rename $v=v_0$, $w = v_1$, then add the sequence of functions $v_k$, $k\ge 2$, and  consider the infinite system
\begin{equation}
\left\{
\begin{array}{lcl}
v_{k,t} &=& u\,v_{k,x},
\\
v_{k,y} &=&v_{k+1} -v_{k,x}\, \ln u_x.
\end{array}
\qquad k \ge 0.
\right.
\label{closed_covering_1}
\end{equation}
The compatibility conditions of this system coincides with equation \eqref{first_eq}. System \eqref{closed_covering_1} can be written in the finite form by introducing the function 
\begin{equation}
r=\mathrm{e}^{-\lambda\,y}\,\sum \limits_{k=0}^{\infty} \lambda^k\,v_k
\label{r_series}
\end{equation}
where $\sum \limits_{k=0}^{\infty} \lambda^k\,v_k$ is a formal series with respect to a formal parameter $\lambda$. Then we have
\begin{equation}
\left\{
\begin{array}{lcl}
r_{t} &=& u\,r_{x},
\\
r_{y} &=& -r_x\,\ln u_x.
\end{array}
\right.
\label{closed_covering_2}
\end{equation}
This system provides a Lax representation for equation \eqref{first_eq}.

\vskip 5 pt
Direct computations\footnote[2]{We carried out computations of generators of contact symmetries in the {\it Jets} software      \cite{Jets}.} give the following statement:

\vskip 5 pt
\noindent
{\sc Proposition 1}.  
{\it The contact symmetry algebra of equation \eqref{first_eq} is generated by the vector fields}
\begin{equation}
A_0\,\partial_t
-A_0^{\prime}\,y\,\partial_x  -(A_0^{\prime}\,u-A_0^{\prime\prime}\,y)\,\partial_u,
\quad
A_1\,\partial_t-A_1^{\prime}\,\partial_u,
\quad
x\,\partial_x+y\,\partial_y+u\,\partial_u,
\quad
\partial_y,
\label{symmetries_of_first_eq}
\end{equation}
{\it where $A_i=A_i(t)$ are arbitrary smooth functions of $t$. Restricting these functions to polynomials  gives the Lie algebra
isomorphic to $\mathfrak{a}_{2}\ltimes \mathfrak{q}_{1,-1}$.  The contact symmetry algebra of system \eqref{first_eq}, \eqref{wt} 
is obtained by appending the vector field $B\,\partial_v$ with arbitrary smooth function $B=B(y,v)$ to the vector fields \eqref{symmetries_of_first_eq}.
}
\hfill $\Box$

%===============================================================

\section{3D generalized Hunter--Saxton equation}
\label{section_5}

The Lie algebra $\mathfrak{a}_{0}\ltimes \mathfrak{q}_{1,-1/2}$ admits the outer  derivation 
$D_2 (s^p\,f(t))= s^{p+1}\,f^{\prime\prime}(t)$, that is,  
\[
D_2(s^p\otimes V_k) =
\left\{
\begin{array}{lcl}
k\,(k-1)\,s\otimes V_{k-2}, &~& p = 0\,\,\, \mathrm{and}\,\,\, k \ge2,
\\
0, && p =1 \,\,\, \mathrm{or}\,\,\, k \in \{0, 1\}.
\end{array}
\right.
\]
This derivation produces the right extension 
$\mathfrak{b}_1 \ltimes \mathfrak{q}_{1,-1/2} = \langle Z_1\rangle \ltimes (\mathfrak{a}_{0}\ltimes \mathfrak{q}_{1,-1/2})$ of 
$\mathfrak{a}_{0}\ltimes \mathfrak{q}_{1,-1/2}$, where  $[Z_1,V_k]_{-1/2} = D_2(V_k)$ and  $[Z_0,Z_1]_{-1/2} = Z_1$. 
Denote by $\alpha_1$ the dual form to $-Z_1$, that is, put $\alpha_1(Z_1)=-1$, $\alpha_1(Z_0) =0$, $\alpha_1(V_k)=0$.
Then the structure equations for $\mathfrak{b}_1 \ltimes \mathfrak{q}_{1,-1/2}$ acquire the form 
\begin{equation}
\left\{
\begin{array}{lcl}
d\Theta &=& \displaystyle{
\Theta_{h_1} \wedge \left(\Theta -2\,h_0 \,\Theta_{h_0}\right)
+ h_0\,\alpha_0 \wedge \Theta_{h_0} -  h_0\, \alpha_1 \wedge \Theta_{h_1h_1},
}
\\
d\alpha_0 &=&\displaystyle{0,\phantom{\frac{A}{A}}}
\\
d\alpha_1 &=& \displaystyle{\alpha_0 \wedge \alpha_1.\phantom{\frac{A}{A}}}
\end{array}
\right.
\label{q_1_-2_right_extension}
\end{equation}
The Lie algebra $\mathfrak{b}_1 \ltimes \mathfrak{q}_{1,-1/2}$ admits the sequence of twisted extensions 
$\mathfrak{b}_k \ltimes \mathfrak{q}_{1,-1/2}$, $k\ge 2$, as it was described when constructing system \eqref{hat_qNek_SE}. In this section we need the second extension from this series. Specifically, we have 
$[\alpha_0 \wedge \alpha_1] \in H^2_{\alpha_0}(\mathfrak{b}_1 \ltimes \mathfrak{q}_{1,-1/2})$, hence we append equation 
\begin{equation}
d\alpha_2 = \alpha_0 \wedge \alpha_2+\alpha_0\wedge \alpha_1,
\label{q_1_-2_first_twisted_extension}
\end{equation}
to system \eqref{q_1_-2_right_extension} and get the structure equations of $\mathfrak{b}_2 \ltimes \mathfrak{q}_{1,-1/2}$. Then 
$[\alpha_1 \wedge \alpha_2] \in H^2_{2 \alpha_0}(\mathfrak{b}_2 \ltimes \mathfrak{q}_{1,-1/2})$, and the structure equations of 
$\mathfrak{b}_3 \ltimes \mathfrak{q}_{1,-1/2}$ are obtained by  adding equation
\begin{equation}
d\alpha_3= 2\,\alpha_0 \wedge \alpha_3+\alpha_1\wedge \alpha_2.
\label{q_1_-2_second_twisted_extension}
\end{equation}
to system \eqref{q_1_-2_right_extension}, \eqref{q_1_-2_first_twisted_extension}.

Successively integrating equations \eqref{q_1_-2_right_extension}, \eqref{q_1_-2_first_twisted_extension}, and
\eqref{q_1_-2_second_twisted_extension} by applying Frobenius' theorem,  we get
\[
\fl
\qquad
\alpha_0 = \frac{dq}{q},
\quad
\alpha_1 = q\,dy,
\quad
\alpha_2=q \,(dw+\ln q\, \,dy),
\quad
\alpha_3  =q^2\,(dv-w\,dy),
\]
\[
\fl\qquad
\theta_{0,0}= a\,dt,
\quad
\theta_{0,1} = \frac{da}{a}+p_1\,dt,
\quad
\theta_{0,2} = \frac{dp_1}{a}+p_2\,dt,
\]
\[
\fl
\qquad
\theta_{1,0} = q\,a^{-1}\,(dx +p_1\,dt+u\,dt),
\quad
\theta_{1,1} = q\,a^{-2}\,(du -p_1\,dx +(a\,p_2-p_1^2)\,dy+p_3\,dt),
\]
where $q \neq 0$, $a \neq 0$, $t$, $x$, $y$, $u$, $v$, $w$, $p_1$, $p_2$, $p_3$ are free parameters. By altering notation as 
$p_1 =u_x$, $p_2=a^{-1}\,(u_x^2-u_y)$, and $p_3 = -u_t$  we obtain $\theta_{1,1} = q\,a^{-2}\,(du-u_t\,dt-u_x\,dx-u_y\,dy)$.   Then we consider the linear combination
$\tau = \alpha_3 - \theta_{1,0} = q^2\,\left(dv- a^{-1}\,q^{-1}\,(u\,dt +dx+(a\,q\,\,w+u_x)\,dy)\right)$,
put $a=q^{-1}\,v_x^{-1}$ and obtain
\[
\tau = q^2\,(dv -u\,v_x\,dt - v_x\,dx -(w+u_x\,v_x)\,dy).
\]
Upon restriction to the sections of the bundle $\mathbb{R}^6 \rightarrow \mathbb{R}^3$,
$(t,x,y,u,v,w) \mapsto (t,x,y)$ this form produces the over-determined system
\begin{equation}
\left\{
\begin{array}{lcl}
v_t &=& u\,v_x,
\\
v_y &=&w +u_x\,v_x,
\end{array}
\right.
\label{v_covering_for_HS2D}
\end{equation}
which is compatible by virtue of two equations
\begin{equation}
u_{tx}=u\,u_{xx}-u_x^2-u_y
\label{HS2D}
\end{equation}
and
\begin{equation}
w_t = u\,w_x.
\label{open_covering_2}
\end{equation}
The symmetry reduction of equation \eqref{HS2D} with respect to $u_y=0$  coincides with the generalized Hunter-- Saxton equation
\cite{HunterSaxton1991,HunterZheng1994,BealsSattingerSzmigielski2001,Morozov2021c}
\begin{equation}
u_{tx} = u\,u_{xx} + \beta\,u_x^2
\label{gHS}
\end{equation}
with the special value $\beta = -1$ of parameter $\beta$. Hence \eqref{HS2D} can be considered as a three-dimensional generalization of the particular case  $u_{tx} = u\,u_{xx}-u_x^2$ of equation \eqref{gHS}.

Likewise to Section \ref{section_4}, we can find the Lax representation of equation \eqref{HS2D} by  renaming $v=v_0$, $w=v_1$ 
and including equation \eqref{open_covering_2} in the infinite system 
\begin{equation}
\left\{
\begin{array}{lcl}
v_{k,t} &=& u\,v_{k,x},
\\
v_{k,y} &=&v_{k+1} +u_x\,v_{k,x},
\end{array}
\qquad k \ge 0.
\right.
\label{HS2D_covering_1}
\end{equation}
System \eqref{HS2D_covering_1} is compatible by virtue of equation \eqref{HS2D}. Series \eqref{r_series}   allows one to write \eqref{HS2D_covering_1} in the form
\begin{equation}
\left\{
\begin{array}{lcl}
r_{t} &=& u\,r_{x},
\\
r_{y} &=& u_x\,r_{x}.
\end{array}
\right.
\label{HS2D_covering_2}
\end{equation}
\vskip 5 pt
We have
\vskip 5 pt
\noindent
{\sc Proposition 2}. {\it The contact symmetry algebra of equation \eqref{HS2D} admits generating vector fields}
\[
A_0\,\partial_t+(A_0^{\prime\prime}\,y-A_0^{\prime}\,x)\,\partial_x
-(2\,A_0^{\prime}\,u-A_0^{\prime\prime}\,x+A_0^{\prime\prime\prime}\,y)\,\partial_u,
\quad
A_1\,\partial_x-A_1^{\prime}\,\partial_u,
\]
\[
x\,\partial_x+y\,\partial_y+u\,\partial_u,
\quad 
\partial_y,
\]
{\it where $A_i=A_i(t)$ are arbitrary smooth functions of $t$.  Restriction of these functions to $\mathbb{R}[t]$ gives the Lie algebra
isomorphic to $\mathfrak{b}_2 \ltimes \mathfrak{q}_{1,-1/2}$. The contact symmetry algebra of system \eqref{HS2D}, \eqref{open_covering_2} has additional generating vector field $B\,\partial_v$ with arbitrary smooth function $B=B(y,v)$.
}
\hfill $\Box$

%====================================================================

\section{Integrable equation associated to $\mathfrak{a}_2 \ltimes \mathfrak{q}_{2,-1}$}
\label{section_6}

The structure equations of the Lie algebra $\mathfrak{a}_3 \ltimes \mathfrak{q}_{2,-1}$ have the form  \eqref{hat_q1-12_SE}, 
where now the formal series $\Theta$ is given by formula \eqref{Theta_N_def} with $N=2$. We take the same forms $\alpha_0$, ..., $\alpha_3$,  and $\theta_{0,0}$ as in Section \ref{section_4}, and put 
\[
\theta_{1,0}= q\,(db - \ln a\, dy -p_1\,dt),\quad
\theta_{2,0} = -q^2\,a^{-1}\,(dx+p_1\,dy-p_2\,dt).
\]
Then the linear combination
\[
\fl
\theta_{1,0}+\theta_{2,0}+\alpha_2 =
q\,\left(db+dw
+\frac{a\,p_1+q\,p_2}{a}\,dt
-\frac{q}{a}\,dx
-\frac{a\,(\ln a -\ln q)+q\,p_1}{a}\,dy\right)
\]
after altering notation  $b=u-w$, $q=a\,u_x$,  $p_1=(u_y+\ln u_x)\,u_x^{-1}$, and $p_2 =- (u_t\,u_x+u_y+\ln u_x)\,u_x^{-2}$
acquires the form $\theta_{1,0}+\theta_{2,0}+\alpha_2 =q\,(du-u_t\,dt-u_x\,dx-u_y\,dy)$, while for the form 
\[
\fl
\tau= \alpha_3+\theta_{2,0}=a^2\,u_x^2\, 
\left(
dv-\frac{1}{a}\,\left( dx+\frac{u_t\,u_x+u_y+\ln u_x}{u_x^2}\,dt+\frac{a\,u_x\,w+u_y+\ln u_x}{u_x}\,dy\right)
\right) 
\]
after renaming $a=v_x^{-1}$ we have 
\[
\fl
\tau=\frac{u_x^2}{v_x^2}\, 
\left(
dv-v_x\,\left( dx+\frac{u_t\,u_x+u_y+\ln u_x}{u_x^2}\,dt\right)+\frac{u_x\,w\,+(u_y+\ln u_x)\,v_x}{u_x}\,dy
\right). 
\]
Restricting this onto sections of the bundle $(t,x,y,u,v,w) \mapsto (t,x,y)$ produces system 
\begin{equation}
\left\{
\begin{array}{lcl}
v_t &=& \displaystyle{\frac{u_t\,u_x+u_y+\ln u_x}{u_x^2}\,v_x,\phantom{\frac{\frac{A}{A}}{\frac{A}{A}}}}
\\
v_y &=& \displaystyle{\frac{u_y+\ln u_x}{u_x}\,v_x +w,\phantom{\frac{\frac{A}{A}}{\frac{A}{A}}}}
\end{array}
\right.
\label{v_covering_for_eq_08_04_2021}
\end{equation}
which  is compatible by virtue of equations 
\begin{equation}
u_{yy} = u_{tx}-\frac{(u_y +\ln u_x)^2+u_t\,u_x}{u_x^2}\,u_{xx}+\frac{2\,(u_y+\ln u_x)-1}{u_x}\,u_{xy}
\label{eq_08_04_2021} 
\end{equation}
and 
\begin{equation}
w_t=\frac{u_t\,u_x+u_y+\ln u_x}{u_x^2}\,w_x.
\label{eq_08_04_2021_for_wt}
\end{equation}
The last equation is a copy of the first equation in \eqref{v_covering_for_eq_08_04_2021}. Therefore we can take $w$ as a new pseudopotential, rename    $v=v_0$, $w=v_1$, add the infinite sequence of pseudopotentials $v_j$, $j \ge 2$,  and consider infinite system  
\begin{equation}
\left\{
\begin{array}{lcl}
v_{k,t} &=& \displaystyle{\frac{u_t\,u_x+u_y+\ln u_x}{u_x^2}\,v_{k,x},\phantom{\frac{\frac{A}{A}}{\frac{A}{A}}}}
\\
v_{k,y} &=& \displaystyle{\frac{u_y+\ln u_x}{u_x}\,v_{k,x} +v_{k+1},
\phantom{\frac{\frac{A}{A}}{\frac{A}{A}}}}
\end{array}
\right.
\qquad k \ge 0.
\label{v_k_covering_for_eq_08_04_2021}
\end{equation}
This system is compatible by virtue of  equation \eqref{eq_08_04_2021} and thus defines a Lax re\-pre\-sen\-ta\-ti\-on thereof.  
Introducing series \eqref{r_series}, we rewrite \eqref{v_k_covering_for_eq_08_04_2021} in the form of another Lax representation 
\begin{equation}
\left\{
\begin{array}{lcl}
r_t &=& \displaystyle{\frac{u_t\,u_x+u_y+\ln u_x}{u_x^2}\,r_{x},\phantom{\frac{\frac{A}{A}}{\frac{A}{A}}}}
\\
r_{y} &=& \displaystyle{\frac{u_y+\ln u_x}{u_x}\,r_{x}
\phantom{\frac{\frac{A}{A}}{\frac{A}{A}}}}
\end{array}
\right.
\label{r_covering_for_eq_08_04_2021}
\end{equation}
for equation \eqref{eq_08_04_2021}.

\vskip 5 pt

\noindent
{\sc Proposition 3}. 
{\it The following vector fields}
\[
A_0\,\partial_t
-\left(A_0^{\prime}\,x+\frac{1}{2}\,A_0^{\prime\prime}\,y^2\right)\,\partial_x
-A_0^{\prime}\,y\,\partial_u,
\quad
A_1^{\prime}\,y\,\partial_x+A_1\,\partial_u,
\quad
A_2\,\partial_x,
\]
\[
2\,x\,\partial_x+y\,\partial_y+(u+y)\,\partial_u,
\quad
\partial_y.
\]
{\it 
with arbitrary smooth functions $A_i=A_i(t)$ are  generators for the contact symmetry algebra of equation \eqref{eq_08_04_2021}.
Restricting these functions to polynomials gives the Lie algebra isomorphic to $\mathfrak{a}_2 \ltimes \mathfrak{q}_{2,-1}$.
The contact symmetry algebra of system \eqref{eq_08_04_2021}, \eqref{eq_08_04_2021_for_wt}  is obtained by appending the 
vector field $B\,\partial_v$ with arbitrary smooth function $B=B(y,v)$. 
}
\hfill $\Box$

\section{The degenerate heavenly equation}
\label{section_7}

In this section we construct a Lie algebra that includes the algebra of polynomial Hamiltonian vector fields
\[
\mathfrak{h} = \left\langle W_{m,n} =n\,t^m\,x^{n-1}\,\partial_t-m\,t^{m-1}\,x^n\,\partial_x \,\,
\vert\,\, m, n \in \mathbb{N}_0, \,\,m^2+n^2\neq 0\right\rangle
\]
on $\mathbb{R}^2$ as a subalgebra and has a twisted extension that generates an integrable {\sc pde}. To find such a Lie algebra 
we proceed as follows. The Lie algebra $\mathfrak{h}$   admits the  grading  
$\mathrm{gr}(W_{m,n}) = m+n-2$ that defines the outer derivation 
$D\colon W_{m,n} \mapsto (m+n-2)\,W_{m,n}$ and the right extension
$\langle Z \rangle \ltimes \mathfrak{h}$ with $[Z, W_{m,n}] = D(W_{m,n})$.
Denote the dual forms to $W_{p,q}$ and  $-Z$ as $\theta_{i,j}$ and $\alpha$, so
$\theta_{i,j}(W_{m,n}) = \delta_{i,m}\,\delta_{j,n}$, $\theta_{i,j}(Z)=0$, $\alpha(W_{m,n})= 0$, $\alpha(Z)=-1$.
Further, we take  $\mathfrak{w} = \langle V_k = \frac{1}{k!}\,y^k\,\partial_y\,\,\vert\,\,k \ge 0\rangle$ and consider the direct sum of Lie algebras
\[
\widetilde{\mathfrak{h}} = (\langle Z \rangle \ltimes \mathfrak{h}) \oplus \mathfrak{w}.
\]
Denote by $\omega_m$ the dual forms for $V_k$, so  $\omega_m(V_k) = \delta_{m,k}$, $\omega_m(W_{k,n}) = \omega_m(Z) =0$,
and $\theta_{m,n}(V_k) = \alpha(V_k) =0$. Put
\[
\Theta = \sum \limits_{
m\ge 0,\, n \ge 0,\,
m^2+n^2\neq 0
}
\,
\frac{h_1^m\,h_2^n}{m!\,n!} \,\theta_{m,n},
\qquad
\Omega = \sum \limits_{k \ge 0}
\,
\frac{h_3^k}{k!}\, \omega_{k},
\]
where $h_i$ are formal parameters such that $dh_i=0$.  Then the structure equa\-ti\-ons for the Lie algebra
$\widetilde{\mathfrak{h}}$ acquire the form
\begin{equation}
\left\{
\begin{array}{lcl}
d\Theta &=& - \Theta_{h_1} \wedge \Theta_{h_2}
-\alpha \wedge \left(h_1\,\Theta_{h_1}+h_2\,\Theta_{h_2} - 2\,\Theta\right),
\\
d\Omega &=& \Omega_{h_3}\wedge \Omega,
\\
d\alpha &=& 0.
\end{array}
\right.
\label{check_h_SE}
\end{equation}
This system entails
$H^1(\widetilde{\mathfrak{h}}) = \langle \alpha \rangle$ and
$[ \theta_{1,0} \wedge \theta_{0,1} ] \in H^2_{2\alpha}(\widetilde{\mathfrak{h}})$.
Hence the Lie algebra $\widetilde{\mathfrak{h}}$ admits the twisted extension defined by appending equation
\begin{equation}
d\sigma = 2\,\alpha \wedge \sigma +\theta_{1,0} \wedge \theta_{0,1}
\label{hat_h_SE}
\end{equation}
to system \eqref{check_h_SE}.

Applying Frobenius' theorme and integrating equations \eqref{check_h_SE}, \eqref{hat_h_SE} step by step we obtain
\[
\theta_{1,0} = a_{11}\,dt +a_{12}\,dx,
\quad
\theta_{0,1}=a_{21}\,dt+a_{22}\,dx,
\]
\[
\alpha = \frac{1}{2}\,\frac{dq}{q}, \quad
q = \mathrm{det}\,\left(
\begin{array}{lcl}
a_{11} &~& a_{12}
\\
a_{21} &~& a_{22}
\end{array}
\right),
\]
\[
\theta_{2,0} = \frac{1}{q}\,(a_{12}\,da_{11}-a_{11}\,da_{12})+b_1\,dt+b_2\,dx,
\]
\[
\theta_{0,2} = \frac{1}{q}\,(a_{22}\,da_{21}-a_{21}\,da_{22})+b_3\,dt+b_4\,dx,
\]
\[
\theta_{1,1} = \frac{1}{2\,q}\,(a_{22}\,da_{11}-a_{11}\,da_{22}+a_{12}\,da_{21}-a_{21}\,da_{12})+b_5\,dt+b_6\,dx,
\]
\[
\omega_0 = p\,dy,
\quad
\sigma= q\,(dv-x\,dt),
\]
where $a_{i,j}$, $b_1$, ... , $b_4$, $p$, $v$  are free parameters such that $q \neq 0$, $p \neq 0$,
and
\[
b_5 =\frac{1}{q}\,(-a_{21}\,a_{22}\,b_1+a_{21}^2\,b_2+a_{11}\,a_{12}\,b_3-a_{11}^2\,b_4),
\]
\[
b_6 =\frac{1}{q}\,(-a_{22}^2\,b_1+a_{21}\,a_{22}\,b_2+a_{12}^2\,b_3-a_{11}\,a_{12}\,b_4),
\]
Put
$a_{21} =u\,a_{22}$,
$b_3=-a_{22}^2\,q^{-1}\,u_t$,
$b_4= - a_{22}^2\,q^{-1}\,u_x$,
$p= - a_{22}^2\,q^{-1}\,u_y$.
This yields
$\theta_{0,2}-\omega_0 = \frac{a_{22}^2}{q}\,(du-u_t\,dt-u_x\,dx-u_y\,dy)$.  Consider the linear combination
\[\fl
\tau = \sigma +\theta_{1,0}-\textfrac{1}{2}\,\omega_0
\]
\[\fl
\qquad
= a_{22}\,(a_{11}-u\,a_{12})\,\left(dv+\frac{dx}{a_{11}-u\,a_{12}}
-\frac{(x\,(a_{11}-u\,a_{12}) -u)\,dt}{a_{11}-u\,a_{12}}-\frac{u_y\,dy}{2\,(a_{11}-u\,a_{12})^2}\right).
\]
By altering notation $a_{11} = u\,a_{12}-v_x^{-1}$ we get
\[
\tau = -\frac{a_{22}}{v_x}\,\left(dv-v_x\,dx -(x+u\,v_x)\,dt-\textfrac{1}{2}\,u_y\,v_x^2\,dy\right).
\]
Then upon restricting  form $\tau$ to the sections of the bundle $\mathbb{R}^5 \rightarrow \mathbb{R}^3$,
$(t,x,y,u,v) \mapsto (t,x,y)$, we obtain the over-determined system
\begin{equation}
\left\{
\begin{array}{lcl}
v_t &=& x+u\,v_x,
\\
v_y &=& \frac{1}{2}\,u_y\,v_x^2
\end{array}
\right.
\label{covering_of_eq_5_02_2021}
\end{equation}
The  compatibility condition of this system $(v_t)_y = (v_y)_t$ holds if and only  
\begin{equation}
u_{ty} =u\,u_{xy}-2\,u_x\,u_y.
\label{eq_5_02_2021}
\end{equation}
Therefore 
system \eqref{covering_of_eq_5_02_2021}
defines a Lax representation for equation \eqref{eq_5_02_2021}.

We notice that equation \eqref{eq_5_02_2021} is invariant with respect to translations along $x$, while its Lax representation \eqref{covering_of_eq_5_02_2021} does not admit such translations, cf \cite[Th. 4]{Morozov2009} and   
\cite[Example 7]{Morozov2018}.

Direct computations give 

\vskip 5 pt

\noindent
{\sc Proposition 4}.
{\it
The contact symmetry algebra  of equation \eqref{eq_5_02_2021} has generators
\[
A_x\,\partial_t-A_t\,\partial_x+(A_{tt}-2\,u\,A_{tx}+u^2\,A_{xx})\,\partial_u,
\quad
B\,\partial_y,
\quad
t\,\partial_t+x\,\partial_x,
\]    
where $A=A(t,x)$ and $B=B(y)$ are arbitrary smooth functions of their arguments. Restriction of these functions to polynomials
$A=t^m\,x^n$, $B=y^k$, $m, n, k \in \mathbb{N}_0$, $m^2+n^2 \neq 0$,  produces the Lie algebra isomorphic to $\widetilde{\mathfrak{h}}$.
}
\hfill $\Box$

\section{Concluding remarks}
In the present paper we have used the method of twisted extensions to derive  new integrable {\sc pde}s related to some infinite-dimensional Lie algebras.  In the obtained examples  as well as in some examples in \cite{Morozov2017}--\cite{Morozov2021b} the integrable equations were constructed starting  from certain extensions of the Lie algebras of the Kac--Moody type or the Lie algebra of
Hamiltonian vector fields on $\mathbb{R}^2$.   In examples of Sections 4, 5, and 6 the symmetry algebras of the obtained integrable
systems turn out to be wider than initial infinite-dimensional Lie algebras used in the construction, while the specific form 
of equations \eqref{wt}, \eqref{open_covering_2}, and \eqref{eq_08_04_2021_for_wt} allowed us to find infinite-com\-po\-nent Lax representations \eqref{closed_covering_1},  \eqref{HS2D_covering_1}, and \eqref{v_k_covering_for_eq_08_04_2021}
as well as one-com\-po\-nent Lax re\-pre\-sen\-ta\-ti\-ons  \eqref{closed_covering_2},  \eqref{HS2D_covering_2}, and \eqref{r_covering_for_eq_08_04_2021} for equations \eqref{first_eq}, \eqref{HS2D}, and \eqref{eq_08_04_2021}, respectively.
The polynomial parts of symmetry algebras of  equations \eqref{first_eq}, \eqref{HS2D}, and \eqref{eq_08_04_2021}
coincide with the infinite-dimensional Lie algebras whose twisted extensions were used to construct systems 
\eqref{open_covering_1}, \eqref{v_covering_for_HS2D}, and \eqref{v_covering_for_eq_08_04_2021}.

We hope further examples will enlighten relations between structure theory of Lie algebras and integrable {\sc pde}s. In particular, it 
is important to address the following issues in the future research:
\begin{itemize}
\item
to find other examples of integrable systems related with the Lie algebras of the Kac--Moody type with small values of  $N$, 
\item
to consider extensions of general deformations of the Lie algebras $\mathbb{R}_{N}[s] \otimes \mathfrak{w}$,
cf.  \cite{Zusmanovich2003}, 
\item
to construct other integrable systems whose symmetry algebras are extensions of the Lie algebras of Hamiltonian vector fields,
\item
to generalize the technique used in the present paper on the Lie--Rinehart algebras, cf. \cite{Morozov2021d}.
\end{itemize}

\section*{Acknowledgments}

I am very grateful to I.S. Krasil${}^{\prime}$shchik for useful discussions.
I thank E.V. Ferapontov and M.V. Pavlov for important remarks.

\bibliographystyle{amsplain}

\end{document}